\documentclass[sigconf]{acmart}

\usepackage{xurl}  

\usepackage{graphicx}
\usepackage{amsmath}
\usepackage{hyperref}
\usepackage{multirow, multicol, makecell}

\usepackage{tabularx}
\usepackage{booktabs}
\usepackage{flushend}

\newcommand{\change}[1]{\textcolor{black}{#1}}

\usepackage{xcolor}
\definecolor{cornflowerblue}{HTML}{4D94FF}

\ccsdesc[500]{Social and professional topics}
\ccsdesc[500]{Social and professional topics~Informal education}

\keywords{Novice Programmers, Empirical Study, Learning-Centered Emotions, Affect-Aware Educational Software Support}

\copyrightyear{2026}
\acmYear{2026}
\setcopyright{cc}
\setcctype{by}
\acmConference[ICSE SEET '26]{48 ACM International Conference on the
 Software Engineering}{April 12-18, 2026}{Rio De Janerio, Brazil}

\title{Learning Programming in Informal Spaces: Using Emotion as a Lens to Understand Novice Struggles on r/learnprogramming}

\begin{document}

\author{Alif Al Hasan*}
\affiliation{%
  \institution{Case Western Reserve University}
  \city{Cleveland, Ohio}
  \country{USA}
}
\email{alifal.hasan@case.edu}

\author{Subarna Saha*}
\affiliation{%
  \institution{Jahangirnagar University}
  \city{Dhaka}
  \country{Bangladesh}
}
\email{subarna.stu2019@juniv.edu}

\author{Mia Mohammad Imran}
\affiliation{%
  \institution{Missouri University of Science and Technology}
  \city{Rolla, Missouri}
  \country{USA}
}
\email{imranm@mst.edu}

\thanks{* These authors contributed equally to this work.}

\begin{abstract}
Novice programmers experience emotional difficulties in informal online learning environments, where \textit{Confusion} and \textit{Frustration} can hinder motivation and learning outcomes. This study investigates novice programmers' emotional experiences in informal settings, identifies causes of emotional struggle, and explores design opportunities for affect-aware support systems. We manually annotated 1,500 posts from \textit{r/learnprogramming} using the \textbf{Learning-Centered Emotions} framework, applying clustering, and axial coding. \textit{Confusion}, \textit{Curiosity}, and \textit{Frustration} dominated emotional experiences, sometimes co-occurring and linked to early learning stages. Positive emotions were infrequent. The primary emotional triggers included ambiguous errors, unclear learning pathways, and misaligned resources. 
We identify five key areas where novice programmers need support in informal space: \textit{Stress Relief and Resilient Motivation}, \textit{Topic Explanation and Resource Recommendation}, \textit{Strategic Decision Making and Learning Guidance}, \textit{Technical Support}, and \textit{Supporting Acknowledgment}. 
Our findings underscore the need for intelligent, affect-sensitive mechanism for providing timely support aligned with learners' emotional states.

\end{abstract}

\maketitle


\section{INTRODUCTION}

Learning to program presents significant challenges for novice programmers~\cite{lopez2021putting}, often triggering various emotional responses that shape their learning journey and influence their success~\cite{graesser2020emotions}. 
Novice programmers increasingly turn to informal learning environments like community-driven platforms (\textit{Reddit}, \textit{YouTube}) for help. A 2024 Stack Overflow survey found that 82.1\% of developers use online resources to learn programming, with 41.2\% using social media to learn~\cite{stackoverflow2024survey}. 

\begin{figure}[tb]
\centering
\includegraphics[width=\linewidth]{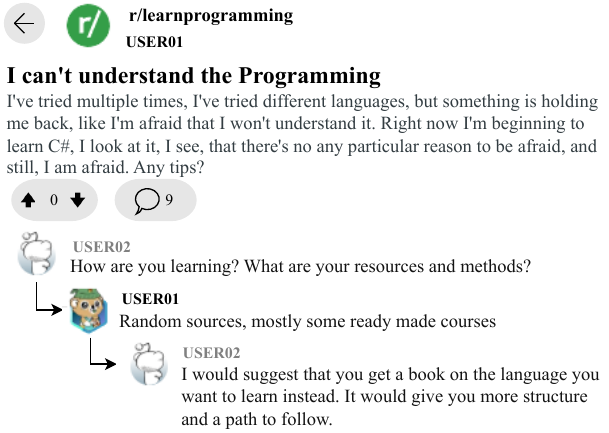}
\caption{Example of concern and community support in \textit{r/learnprogramming}}
\label{fig:motivating-example}
\end{figure}

However, these environments often lack structured guidance, leaving learners responsible for managing their own progress and decision-making. 
In such settings, learners must actively guide their educational experiences by setting personal goals, monitoring their understanding, and adjusting strategies based on feedback~\cite{haythornthwaite2018learning}. Emotional regulation plays a critical role, as emotions such as \textit{Confusion}, \textit{Frustration}, \textit{Curiosity}, and \textit{Anxiety} significantly impact engagement and persistence~\cite{calvo2011new}. For novices, balancing both the cognitive and emotional demands of learning can be especially difficult due to limited experience in navigating complex learning situations~\cite{rogerson2010fear}. Unlike institutional settings with predefined support structures, informal platforms depend on inconsistent, community-driven assistance~\cite{zheng2022exploratory}.

While existing studies have explored emotional experiences in formal education~\cite{reicher2010building, bosch2017affective, humphrey2007emotional} and dedicated software development focused platforms like Stack Overflow and GitHub~\cite{calefato2018ask, imran2022data}, \change{emotional dynamics in informal online learning communities, such as, Reddit, YouTube, Twitch, etc, remain under-examined.
Among these, Reddit, and specifically the \textit{r/learnprogramming} subreddit, offers a uniquely valuable site for such investigation. 
Unlike Stack Overflow, which encourages a technical, Q\&A structure~\cite{calefato2018ask}, Reddit often encourage anonymous, open-ended discussion where users may share both emotional struggles and cognitive difficulties at the same time~\cite{haythornthwaite2018learning, alghamdi2024thematic}. An example of this type of interaction is shown in Figure~\label{fig:motivating-example}, where a novice programmer expresses his emotional struggles stemming from uncertainty about how to begin learning.}

This paper examines the emotional experiences of novice programmers in informal online settings by analyzing 1,500 posts from Reddit’s \textit{r/learnprogramming} subreddit. Drawing on the Learning-Centered Emotions (LCE) framework~\cite{craig2004affect}, we identify emotional states and their causes to inform the design of intelligent support mechanisms. Specifically, we address the following RQs:

{\em RQ1: What are the most prevalent Learning-Centered Emotions experienced by novice programmers in informal learning environments?}

→ This RQ examines the emotional states most commonly experienced by novice learners. While well-known models like Ekman’s Basic Emotions~\cite{ekman}, which identifies six universal emotions (e.g., anger, fear, sadness), Shaver’s Hierarchical Model~\cite{Shaver}, and Plutchik’s Wheel of Emotions~\cite{PLUTCHIK19803} provide broad emotion categories, they are not tailored to the dynamics of learning. In contrast, the LCE framework~\cite{craig2004affect} captures emotions that are specific to learning processes, including both cognitive-affective and transitional states. The LCE model defines eight key emotions: \textit{Frustration}, \textit{Confusion}, \textit{Boredom}, \textit{Engagement/flow}, \textit{Curiosity}, \textit{Anxiety}, \textit{Delight}, and \textit{Surprise}. These transitional emotions, particularly \textit{Confusion} and \textit{Curiosity}, are critical in self-directed programming and are notably absent from other models. As Imran et al.~\cite{imran2022data} argue, such categories are especially relevant in software engineering contexts. LCE is also well-suited for analyzing Reddit posts, where learners describe evolving emotional states during problem-solving. Prior work~\cite{calvo2011new, bosch2017affective, d2014confusion} supports its use in adaptive, emotion-aware learning systems.
In our analysis, \textit{Confusion}, \textit{Curiosity} and \textit{Frustration} were the most common and often co-occurred. Positive emotions like \textit{Delight} and \textit{Engagement/flow} were rare but signaled key moments of progress.

{\em RQ2: What are the primary causes of Learning-Centered Emotions experienced by novice programmers?}

→ This RQ uncovers key individual causes that lead to the emotions. \change{We applied DBSCAN clustering~\cite{schubert2017dbscan} to find the common themes and focus on top 5 groups.} We found that debugging challenges, resource mismatches, lack of direction, and conceptual barriers were major causes.

{\em RQ3: What types of support do novice programmers need in the r/learnprogramming community?}

→ Building on the findings of RQ1-RQ2, this RQ examines the specific support needs that novice programmers express or imply in their posts, using open coding and axial coding~\cite{charmaz2006constructing}, we identify five actionable categories that could address both affective and cognitive dimensions of their learning struggles. They are: \textit{Stress Relief and Resilient Motivation}, \textit{Topic Explanation and Resource Recommendation}, \textit{Strategic Decision Making and Learning Guidance}, \textit{Technical Support}, and \textit{Supporting Acknowledgment}.

\paragraph{Contributions.}
This paper contributes by providing an empirical account of novice programmers’ emotions in informal learning environments, offering insight into how affect shapes participation in community-driven spaces like Reddit. 
We present a novel dataset of 1,500 annotated posts labeled for learning-centered emotions (LCEs) and support needs, and introduce a grounded category framework for automated support in informal programming learning.
Our analysis advances understanding of the connections between emotional struggles and expressed support needs, grounding these insights in naturally occurring learner discourse.
Finally, we propose design implications for designing intelligent affect-aware support system, highlighting opportunities for lightweight, scalable interventions, such as community-integrated bots, that can provide timely emotional and cognitive support in informal settings.


{\color{red} Note: All examples cited in the paper are taken directly from \textit{r/learnprogramming} posts and may include spelling or grammatical errors as they appear in the original user submissions.}
\color{black} Our study's datasets, scripts, and annotation instructions are publicly available online at URL: \color{blue}{\url{https://zenodo.org/records/17145615}}\color{black}.

\section{BACKGROUND AND RELATED WORK}

We discuss three key strands of literature relevant to our work: learning in informal space, emotional experiences of novice programmers, and support systems for programming education.

\smallskip
\noindent \textbf{Learning in Informal Space.}
Many learners pursue programming via informal, self-directed means, particularly online \cite{chaudhury2022there, beddie2010informal}. While such platforms expand access by lowering entry barriers and offering flexible resources, they often lack the structured support systems necessary to sustain engagement, especially for novices \cite{kittel2023its}. As a result, beginners frequently encounter challenges such as unclear learning goals, feelings of isolation, and diminished motivation \cite{chaudhury2022there, alghamdi2023exploring}. Social media platforms, in particular, have emerged as common spaces where learners seek both technical assistance and peer encouragement \cite{alghamdi2023exploring}. However, their affordances can also fragment attention and disrupt continuity in learning, as constant exposure to multiple streams of content competes with sustained cognitive effort \cite{want2022comprehensively, lyngs2019self}. In such environments, the ability to self-regulate, by setting personal goals, monitoring one’s progress, and reflecting on outcomes, becomes especially critical. Yet, despite its recognized importance, the role of self-regulation in informal learning remains underexplored compared to formal educational settings \cite{prather2020what, saqr2024mapping}.



\smallskip
\noindent \textbf{Emotional Experiences of Novice Programmers.} 
Emotions play a pivotal role in learning by shaping attention, memory, reasoning, and problem-solving~\cite{tyng2017the, pekrun2014emotions, reichert2024experience}. In programming, the abstract nature of concepts and the frequency of errors make learners particularly vulnerable to negative emotions such as \textit{Frustration}, \textit{Confusion}, and \textit{Anxiety}, which can erode confidence and persistence. At the same time, positive emotions and effective emotion regulation have been shown to foster motivation, sustain engagement, and strengthen self-regulation, creating conditions for more productive learning experiences~\cite{arguedas2016analyzing, tan2021the}. Recognizing this dual influence, research in computing education has increasingly examined how emotions intersect with the learning process. For example, Bosch et al.~\cite{bosch2013emotions} demonstrated that emotional trajectories in novice Python learners closely align with their behavioral responses and performance outcomes, underscoring the link between affective states and observable learning behaviors. Complementing this, Coto et al.~\cite{coto2022emotions} systematically mapped research on emotions in programming education, identifying central variables, methodological approaches, and adaptive strategies that characterize the field. Building on this foundation, more recent affective computing approaches have sought to automatically detect learner emotions in real time. These studies leverage multimodal signals—such as facial expressions, keystroke dynamics, and interaction logs—to identify states like \textit{Engagement}, \textit{Confusion}, and \textit{Frustration}, opening avenues for timely and adaptive support in educational contexts~\cite{lee2019analysis, liu2018predicting, zhuang2022towards}.

\smallskip
\noindent \textbf{Support Systems for Programming Education.} 
Programming education has steadily evolved with tools aimed at improving learning outcomes. Early efforts focused on intelligent tutoring systems that provided adaptive feedback based on learners’ progress and misconceptions~\cite{crow2018intelligent, price2017isnap, weltman2019evaluation}, while more recent approaches use large-scale AI models like GPT-4 to deliver scalable, context-sensitive support~\cite{phung2023generative, jonsson2022cracking}. A growing area of research centers on \textit{intelligent affect detection systems}, which leverage advances in affective computing to detect and respond to learners’ emotional states using multimodal signals such as facial expressions, keystrokes, physiological data, and text sentiment~\cite{lee2019analysis, liu2018predicting, zhuang2022towards}. These affect-aware systems have been shown to improve persistence and motivation by aligning feedback with emotional trajectories~\cite{bosch2017affective, calvo2011new, reyes2021automatic}, underscoring the importance of addressing both cognitive and emotional aspects of learning.

Personalized instruction has also advanced through behavioral and physiological cues that adapt content to individual learners~\cite{tseng2023personalized, karamimehr2023personalised, ismail2025interplay}. Automated feedback tools further reduce instructor workload while offering timely support~\cite{souza2016systematic}. The integration of affect detection into these systems reflects the growing recognition that novice learners need help not just with problem-solving but also with managing frustration, confusion, and motivation.

Unlike prior work focused on formal settings or cognitive support, our study examines the emotional challenges novices face in informal, community-driven environments. By analyzing naturally occurring discourse on Reddit, we capture how learners articulate their struggles and motivations, uncovering design opportunities for affect-aware support grounded in real-world practice.

\begin{figure*}[tb]
    \centering
    \includegraphics[width=\textwidth]{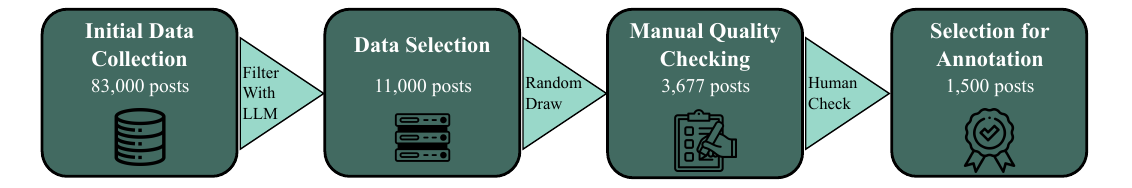}
    \caption{Overview of our r/learnprogramming data collection pipeline.}
    \label{fig:data_collection}
\end{figure*}

\section{METHODOLOGY}\label{dataset}

In this section, we present the methodological steps used in our study. We define the emotion categories based on the Learning-Centered Emotions (LCE) framework, describe our data collection and filtering process, and explain our annotation procedures for identifying emotional states and support needs in posts from novice programmers.

\begin{table*}[tb]
    \centering
    \caption{Learning-Centered Emotions framework: definitions, examples from \textit{r/learnprogramming}, and the inferred causes}
    \label{tab:learning-centered-emotions}
    \begin{tabular}{|l|p{0.28\linewidth}|p{0.37\linewidth}|p{0.18\linewidth}|}
        \hline
        & Definition & Example (from \textit{r/learnprogramming}) & Cause \\
        \hline

        Anxiety & A state of high arousal and negative emotions stemming from fear or worry about potential failure or challenges in learning & 
        \textit{``i have been trying to learn javascript on and off for the past year and i just cannot get it to stick. its causing me so much anxiety that i feel like a complete failure. [...] is there a way past this or to change your mindset? i do not want to give up"} 
        & Feeling a sense of failure and worrying about inability to grasp JavaScript \\
        \hline

        Boredom & A state characterized by low arousal and negative emotions where learners feel disengaged or uninterested in the learning material & 
        \textit{``I'm tired of learning through watching. Any advice? I'm getting kind of tired of watching YouTube videos about learning Python. As much as I'm trying to push through, I feel bored and impatient. [...] I'm a newbie btw in programming and python."}
        & Tired of learning python through watching videos \\
        \hline

        Confusion & A state of moderate arousal where learners experience uncertainty or a lack of clarity in comprehending the material & 
        \textit{``Ask for debugging help : May be some issue in the while loop. Hi everyone. I am using cpp. I would like to seek help for my function [...] It doesn't show the correct value. I think something wrong with my loop but I am not sure where is it. Thanks in advance!"}
        & Unsurety of what is causing the issue in the loop \\
        \hline

        Curiosity & A state of positive arousal and interest where learners are motivated to explore and seek new knowledge & 
        \textit{``Help? I have just learning the basics of C and how everything works. I want to know what should be the next step from here on out? I was thinking about learning c++ from tomorrow, any advice is appreciated." }
        & Eagerness to learn and seeking guidance \\
        \hline

        Delight &  A state of high arousal and positive emotions resulting from the joy or satisfaction of overcoming challenges and successfully grasping the material  & \textit{``If I can program my own sudoku solver from scratch, would that be enough to be considered a "good" programmer? I've been learning python and trying to write a sudoku solver from scratch. I've finally managed to do it and I'm a bit proud of myself. [...]"} & Satisfied at completing the Sudoku solver \\
        \hline

        \multirowcell{2}{Engagement\\/flow} & A state of high concentration and positive valence where learners are fully immersed in their task and find the challenge both stimulating and manageable & \textit{``Advice in python. Hi there. I started learning python not long ago, was reading a book about python for some time, know most of basic stuff. 
        [...] hope some advice where to go what to do to expand and learn as now i am little stuck with no ideas"} & Enjoying learning python and understanding new stuffs \\
        \hline

        Frustration & A state of high arousal and negative emotions where learners feel blocked or overwhelmed by the task & \textit{``i am extremely bad at for loops and i want to improve but i cannot find any good source or site to practice them."} & Bad at for loops and struggling to find resources \\
        \hline

        Surprise & A state of high arousal and mixed valence resulting from an unexpected discovery or realization during the learning process & 
        \textit{``Just learnt Java today. Got a huge shock when I compared it to C++. So usually when I learn a new language, [...] 
        I thought that C++ being a relatively low level language should outperform Java as it's considered a high level language. But apparently not?? [...]"}
        & Surprised when compared the performance of Java to C++ \\
        \hline
        
    \end{tabular}
\end{table*}


\subsection{Emotion Categories}
As mentioned earlier, to systematically capture the affective experiences of novice programmers, we adopt the \textit{Learning-Centered Emotions} (LCE) framework~\cite{craig2004affect}, which identifies emotions that are directly tied to the learning process. \change{This framework has been shown useful in capturing
the dynamic nature of emotions in learning environments, has been adopted in computing research, and has potential uses in educational software development~\cite{yadannavar2024multimodal, gonzalez2021automatic, reyes2021automatic, ecker2023universal, bremner2021learner}.}

The LCE framework encompasses eight key emotional states: \textit{Frustration}, \textit{Boredom}, \textit{Engagement/flow}, \textit{Curiosity}, \textit{Anxiety}, \textit{Delight}, \textit{Confusion}, and \textit{Surprise}.
This framework accounts both negative emotions that hinder learning (e.g., \textit{Frustration}, \textit{Anxiety}, \textit{Boredom}) and positive emotions that facilitate motivation and persistence (e.g., \textit{Delight}, \textit{Engagement}).
It also captures transitional states such as \textit{Confusion} and \textit{Curiosity}, which may lead to either disengagement or deeper understanding depending on how learners cope with uncertainty~\cite{calvo2011new, craig2004affect}. Table~\ref{tab:learning-centered-emotions} provides overview and examples. All examples in this paper are drawn directly from our dataset and are presented exactly as they originally appeared on \textit{r/learnprogramming}.

\begin{figure*}[tb]
\centering
\includegraphics[width=0.7\linewidth]{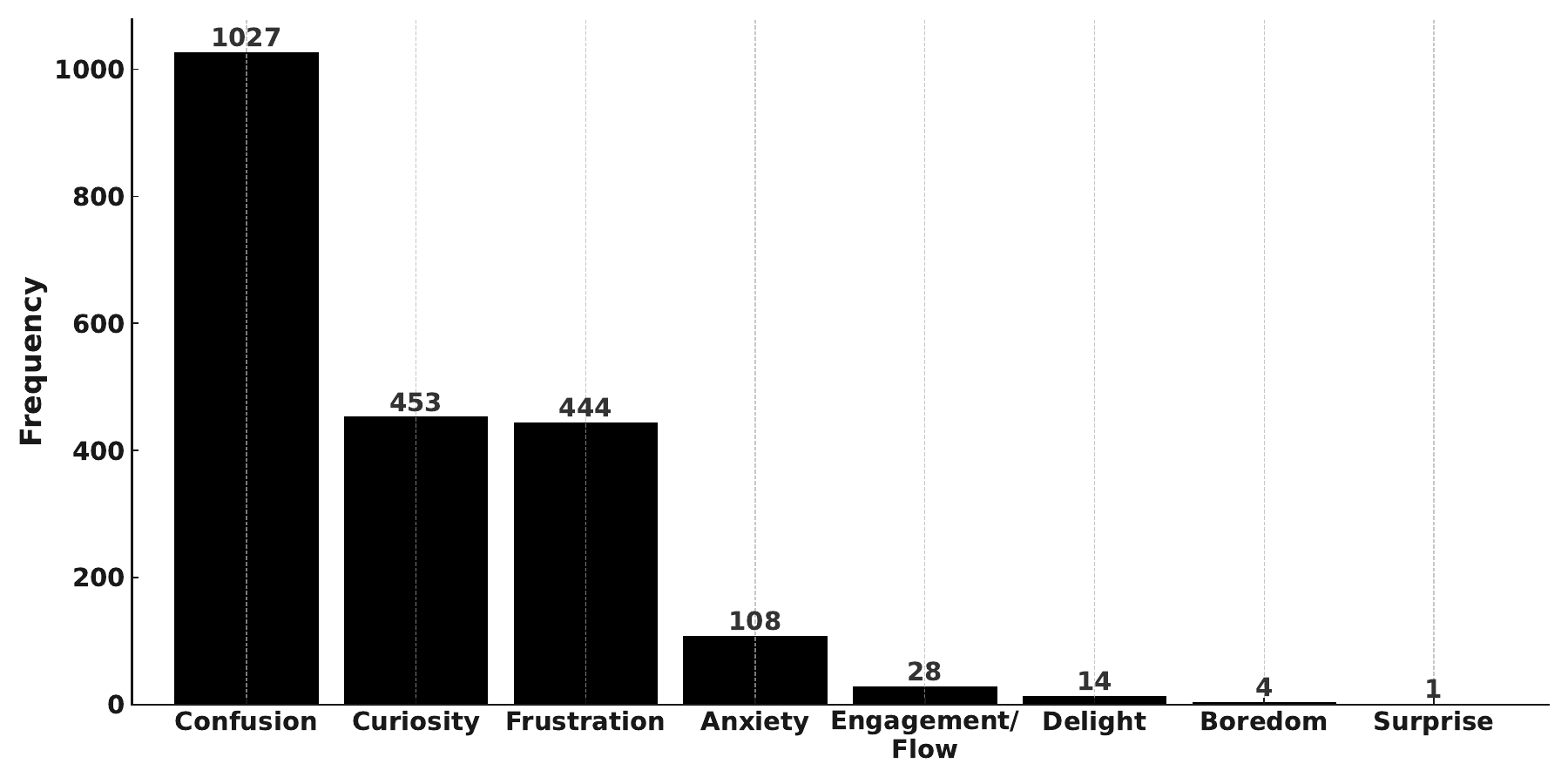}
\caption{Frequency of Learning-Centered Emotions across 1,500 posts.}
\label{fig:count-lce}
\end{figure*}

\subsection{Data Collection} 
Reddit recently gained attention for its role in understanding informal learning communities~\cite{klemmer2024using, da2022slim, hudgins2020informal, alghamdi2024thematic}.
We started by identifying subreddits related to programming, with a particular emphasis on those designed to support beginners. From an initial group of top 30 subreddits by activity as listed by Reddit (communities tab on URL: www.reddit.com/t/programming/), we selected subreddits that explicitly mentioned that they are beginner friendly in their about/wiki section: \textit{r/learnprogramming} and \textit{r/learnpython}. We did not check other learning related subreddits (e.g., \textit{r/learnjavascript}) as they were not listed here.

From the selected subreddit, we sampled default top 100 recent posts. We particularly checked posts in which users specifically identified themselves as novices, beginners, or newcomers to programming. Afterwards, we chose to concentrate just on \textit{r/learnprogramming} because it had the highest number of posts from novice learners (i.e., where the user mentioned they are new to programming using terms such as `new', `beginner', `newbie', `novice', etc). \change{We also observed that very few posts exhibited emotions, which is consistent with findings from previous research in software engineering~\cite{imran2022data, murgia2014developers}.}

We collected all posts from \textit{r/learnprogramming} from January 2023 to November 2024 (83k posts) using Reddit data dump. As this dataset is huge and \change{since 1) not all of them are written by novice programmers, and 2) most of posts do not exhibit emotions, we filtered out the dataset using a model-in-the-loop approach by following Imran et al.~\cite{imran2022data}'s strategy. In their emotional annotation of GitHub comments, Imran et al. first filtered non-neutral data by applying a sentiment analysis tool and afterwards annotated emotions in GitHub comments.} 

\change{We choose to use model-in-the-loop approach with Llama since recent research shows that such a model-in-the-loop annotation methodology works well for this type of data~\cite{sanyal2024machines, zendel2024enhancing, nguyen2024human, wang2024human, zhu2024apt}, including software engineering domain~\cite{imran2025silent}.
We used an LLM with prompt-based filtering to identify posts that contain emotional expressions and are written by novice programmers.
More specifically, we asked the model the following questions using few-shot prompting: a) does the post exhibit any of the following emotions (\textit{Frustration}, \textit{Confusion}, \textit{Boredom}, \textit{Engagement/flow}, \textit{Curiosity}, \textit{Anxiety}, \textit{Delight}, or \textit{Surprise}), and b) is it written by a novice programmer. We used Llama (Llama 3.1-70B version \footnote{{\url{https://ai.meta.com/blog/meta-llama-3-1/}}}) since it was one of the most powerful open-weight models at the time of this experiment.}

We selected posts where the model answered positively to both questions, resulting in 11k posts. \change{We did not select any posts from the remaining 72k instances that are discarded by the LLM model. This aligns with the previous research practices in curating emotion datasets as noted by Koufakou et al.~\cite{koufakou2025review}.}
Since there can be errors by LLM selected posts due to various reasons such as hallucination, misunderstood context, etc, to further ensure relevance to self-learning beginners and emotional content, three authors randomly drew posts one by one from this set and verified each post's suitability. The dataset was shuffled before the random selection procedure. They continued this verification process until reaching 1,500 relevant posts, reviewing a total of 3,677 posts during this procedure. 
Given the labor-intensive nature of manual emotion annotation and the necessity of mitigating annotator subjectivity~\cite{imran2022data}, we capped the dataset at 1,500 posts, aligning with sample sizes used in recent studies on emotional text corpora~\cite{koufakou2025review}. Importantly, this filtered set excludes posts labeled as expressing \textit{neutral} emotions. As noted by Koufakou et al.~\cite{koufakou2025review}, neutral content typically dominates emotion datasets. Therefore, our curated collection of 1,500 posts containing only non-neutral emotional expressions represents a focused and appropriately sized sample for this analysis. An overview of the data collection process
is shown in Figure~\ref{fig:data_collection}.

\subsection{Data Annotation: LCE}

To systematically capture the LCEs present in our dataset, we adopted a coding scheme grounded in Table~\ref{tab:learning-centered-emotions} and informed by prior work on LCEs~\cite{calvo2011new, pekrun2002academic}. Two authors jointly developed and refined an annotation manual, which contained instructions, definitions (e.g., \textit{Frustration}, \textit{Anxiety}), and representative indicators (e.g., misunderstanding a concept, debugging hurdles). As some posts exhibited multiple emotions, we allow multi-label annotations. \change{We followed the annotation procedure by Imran et al. where two annotators initially annotated a joint set, and after achieving a high interrater agreement,the rest were annotated separately by the two annotators~\cite{imran2022data}.} 

Two annotators (authors of this paper) received a shared set of 500 \textit{r/learnprogramming} posts along with the following instructions: 

\begin{quote}
{\em “Use the coding schema outlined in Table~\ref{tab:learning-centered-emotions} to analyze each Reddit post's explicit and implicit emotional content. Identify any Learning-Centered Emotions (LCEs) - Confusion, Frustration, Boredom, Engagement/flow, Curiosity, Anxiety, Delight, or Surprise - and record them in the Emotion column. In the Cause column, explain what triggered each emotion, such as conceptual difficulties, debugging challenges, or progress assessment. Base your annotations on specific textual evidence, supporting any interpretations of implied emotions with brief reasoning.”}
\end{quote}

A detailed version of our annotation scheme can be found in the replication package. The sample size of 500 was selected to ensure high confidence in computing inter-rater agreement metrics~\cite{bujang2017simplified}. Each annotator independently labeled the posts. We calculated Cohen’s Kappa for the eight LCEs. It yielded an agreement score of 0.66, which is considered substantial (\(> 0.6\))~\cite{stemler2019comparison}. The two annotators conducted several meetings, discussed and resolved any disagreements they had. Afterwards, the annotators separately annotated 500 instances each to reach a total count of 1,500 posts~\cite{imran2022data}. Of these, 526 posts contained multiple emotions.

\subsection{Data Annotation: Support Needs}
Since most posts on \textit{r/learnprogramming} involve some form of help-seeking, we extended our analysis to identify the specific types of support that novice programmers either explicitly request or implicitly signal. Our goal was to inform the design of an automated intelligent support systems capable of addressing both emotional and cognitive challenges faced by learners in informal settings.

We followed a qualitative coding approach grounded in open and axial coding~\cite{charmaz2006constructing}. Annotators were instructed to propose up to two plausible automated support categories per post, such as emotional validation, technical debugging guidance, or learning pathway suggestions. Each proposed response was:
\begin{itemize}
    \item Grounded in the post’s emotional tone and content,
    \item Feasible for automation (e.g., chatbot, recommender),
    \item Justified with brief reasoning and textual evidence.
\end{itemize}

The detailed instruction is included in the replication package.
In the \textit{open coding} phase, two annotators independently analyzed 500 posts and derived initial support categories inductively. Through \textit{axial coding}, similar codes were merged and refined, resulting in five stable support need categories. These categories were confirmed to reach thematic saturation during iterative discussions. 
To assess annotation reliability, we computed \textit{Cohen’s Kappa} on posts with both agreement and disagreement cases recorded. The resulting inter-rater agreement was $\kappa = 0.91$, indicating substantial agreement~\cite{stemler2019comparison}. The disagreements were resolved through discussion before proceeding with independent annotations.

Each annotator, afterwards, independently annotated 500 additional posts. The full dataset thus included 1,500 annotated posts. Since posts could have multiple support needs, we found 473 posts included multiple categories, resulting in 1,973 support labels across the dataset.

\subsection{Ethical Considerations}
All data were collected from the publicly accessible \textit{r/learnprogramming} subreddit. No usernames or personal identifiers were recorded, and all quoted content has been anonymized. This study complies with Reddit’s terms of service and follows established ethical guidelines for public online data research~\cite{vitak2016ethics}.

\section{RQ1: PREVALENT EMOTIONS}

{\em RQ1: What are the most prevalent Learning-Centered Emotions experienced by novice programmers in informal learning environments?}

\subsection{Annotation}

To address this, we manually annotated 1,500 posts from the \textit{r/learnprogramming} subreddit using the LCE framework (see Section~\ref{dataset}). The posts were labeled with one or more LCE categories based on both explicit language and inferred emotional cues. \change{Below we note our observations.}

\subsection{Findings}
\noindent
\underline{\textbf {Emotion Distribution}}:
Figure~\ref{fig:count-lce} illustrates the distribution of LCEs in our annotated dataset of novice programmers' experienced emotions in informal online learning communities. \textit{Confusion} is the most prevalent emotion, with 1,027 instances, followed by Curiosity (453) \textit{Frustration} (444), and \textit{Anxiety} (108). These four dominate the emotional state of novice programmers in \textit{r/learnprogramming}, reflecting their significant struggles and inquisitiveness as they navigate programming concepts and tasks. For example:
\textit{“Trying to learn C and I am stuck. I am trying to learn C , I easily get distracted and bored so instead of following tutorials I asked chat-gpt to give me 50 problems which align with the level of knowledge I have about C but I have quickly realized I am struggling to develop solutions, for example a basic program of the 50 was to take a number from user and reverse it even though I initially was on the correct path but I was not able to figure out the entire logic of the program by myself How can I improve? Anyone can google the syntax of languages I want to develop solutions to problems I am given instead of just googling it”} 
This example highlights a mixture of cognitive struggle and intrinsic motivation. The learner expresses not only \textit{Confusion} and \textit{Frustration} but also a strong desire for deeper comprehension—underscoring the complex emotional terrain of learning to program.

\begin{figure}[tb]
    \centering
    \includegraphics[width=1.0\linewidth]{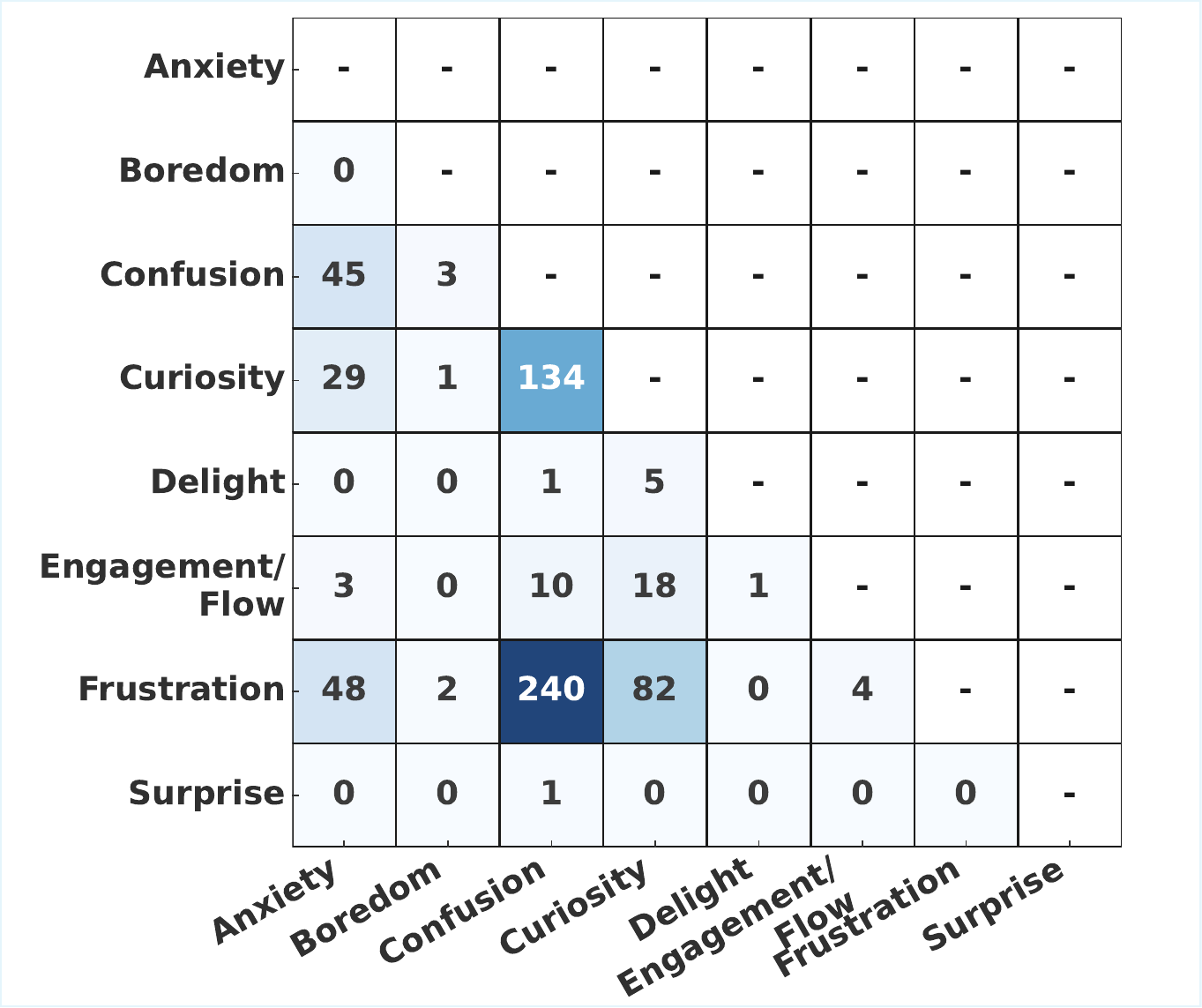}
    \caption{Co-occurrence Matrix of Emotions annotated with multiple LCEs.}
    \label{fig:co-occurence}
\end{figure}

\smallskip
\noindent
\underline{\textbf{Co-Occurrence of Emotions}}: 
Out of 1,500 uniquely annotated posts, 526 were assigned multiple emotional labels, indicating that nearly 35\% of posts reflected a compound emotional state.
\change{Figure~\ref{fig:co-occurence} visualizes the co-occurrence structure, where \textit{Confusion} and \textit{Frustration} form the most frequent emotional dyad.} In these instances, confusion arising from unfamiliar syntax, logical gaps, or debugging issues often escalates into frustration when learners encounter repeated failure or insufficient feedback. For example:
\textit{“Struggling with Python, HTML in a Flask project. I'm a novice coder. I built out a chatbot that uses semantic search and openAI over a few documents of mine related to accessibility. The backend works fine in terms of getting the right messages. But this one issue is driving me nuts. I am trying to render the user message first and then the bot response. But each time the user sends a message, the bot message AND the user message render at the same time. They are in the correct order, but it doesn't feel like a conversational chat bot this way. I dont think I am understanding routes or jjnja templating. [...] Any code that can easily remedy this would be hugely helpful. Sorry for the noob question.”}.
This post shows how technical issues with UI rendering evolved into emotional friction. The mention of “driving me nuts” signals emotional overload, while the apology at the end reflects diminished confidence. These co-occurrence patterns underscore how unresolved \textit{Confusion} can transition into \textit{Frustration}, particularly in environments where learners lack immediate feedback or social support. Understanding these transitions is essential for designing interventions that detect early signs of emotional overload and redirect learners toward scaffolding or peer assistance.

\smallskip
\noindent
\underline{\textbf{Positive Emotions Are Rare}}:
Although positive emotions such as Delight and Engagement/Flow were rare, with only 14 and 28 instances respectively, they offer valuable insight into the conditions that support productive learning. Delight often appeared when learners overcame a persistent challenge, typically accompanied by a sense of personal accomplishment. For example: \textit{“Made my first game loop - day 4 learning python. Feeling really proud of myself Feeling really proud of myself, I started learning python 4 days ago (I have no coding background) and I completed a simple space race gambling simulation gameplay loop inspired by a mini-game I remember from my favourite childhood game “Escape Velocity”. I love game designing and have some experience with it but I have always been terrified of learning to code, so this has been a big boost to my confidence and self-esteem. I can’t even begin to imagine how you all must feel when you finish real projects that you’ve poured months/years of your life into.”}
This indicates that sustained effort and successful task completion contribute to positive emotional outcomes. 
 Engagement or flow, meanwhile, was observed in contexts where learners experienced clarity of goals, a manageable level of difficulty, and a sense of steady progress. 
One post reads: \textit{“I spent hours building a to-do list app. Everything just clicked.”} These moments reflect immersion and motivation. 
While infrequent, such emotions point to potential levers for promoting persistence and satisfaction in learning.

We note that the scarcity of positive emotions may reflect the help-seeking nature of \textit{r/learnprogramming}, where users typically post problems rather than successes. Nevertheless, positive emotion posts reveal user motivations and feelings of belonging~\cite{peacock2019promoting}.

\section{RQ2: EMOTION CAUSES}

\textit{RQ2:  What are the primary causes of Learning-Centered Emotions experienced by novice programmers?}

\subsection{Annotation}

To answer this RQ, we manually annotated the underlying causes of emotions in the same 1,500 posts from the \textit{r/learnprogramming} subreddit (as describe in Section~\ref{dataset}). 
\change{The aim was to move beyond labeling emotions to uncovering the triggers that gave rise to them. To achieve this, we applied clustering techniques to the annotated dataset in order to surface recurring themes in learners’ emotional expressions.} 

\subsection{Clustering}
We selected DBSCAN clustering because of its ability to automatically determine the number of clusters and handle outliers~\cite{schubert2017dbscan}, making it well suited for exploring unknown patterns in emotion triggers. This approach has been shown to be effective in emotion cause analysis~\cite{imran2024uncovering}. DBSCAN requires two key parameters: $\epsilon$, the maximum distance between points in a cluster, and $MinPts$, the minimum number of points required to form a cluster. We conducted manual parameter sweeping to identify optimal values. Initially, we performed standard text preprocessing (removing punctuation, filtering URLs, and lemmatization). Following this, we selected $\epsilon = 0.15$ and $MinPts = 4$, which produced 13 clusters. We then applied cosine similarity on embeddings generated from the \textit{all-mpnet-base-v2} model~\cite{reimers2019sentence}.

\subsection{Findings}

To focus on the most relevant causes, we analyzed the top 5 clusters by size. We utilized \textit{thematic analysis} to identify recurring themes~\cite{maguire2017doing}. Initially, one of the authors conducted a close reading of each post, focusing on the emotional triggers expressed by learners and generated a preliminary set of codes. Then, another author reviewed the codes, and subsequently resolved any discrepancies through discussion to finalize the codes. Each cluster theme is described below:

\smallskip
\noindent \underline{\textbf{Unclear Coding Issues and Debugging Struggles}} (74 instances): Represents confusion, uncertainty, or struggle of the users to find or understand what caused problems in their code, even when errors were clear or hidden. \change{Generally the codes are provided in the posts.} For example:
\textit{“I have created code for a receipt calculator in the pass and code that used parameters that when you did not type letters it would not let you progress to the next input and then same thing with numbers. [...]
whenever I type a number it makes it seem like I didn't type one and repeatedly makes me try to put in another input and won't let me move past it. I created the pattern to only accept numbers 0-9. Any help is appreciated. This is my code: <CODE> [...]”}
\change{This illustrates how novices may perceive even small logical or syntactic issues as overwhelming, which may escalate into frustration.}

\smallskip
\noindent \underline{\textbf{Uncertainty in Programming Kickoff}} (25 instances): Reflects the difficulty beginners faced in figuring out where and how to start learning programming. For example:
\textit{“Where can I learn programming if I'm starting from absolute scratch? As in literally zero knowledge about it. It can be anything. YouTube channel, online course, a website, etc
What app/software can I use to practice programming?
I am learning from absolute scratch, I literally have zero knowledge about programming. [...]
Thank you in advance! I know it's embarrassing to say that I know nothing about programming, but hey, everyone starts as a beginner. [...]”}
\change{Such posts highlight that emotional struggles frequently stem not from coding itself, but from uncertainty in navigating the vast resources and entry points.}

\smallskip
\noindent \underline{\textbf{Trouble with learning new \change{programming languages}}} (18 instances):  \change{This group captures challenges encountered when learners begin working with entirely unfamiliar programming languages. Unlike general debugging struggles, these posts reflect difficulties rooted in understanding language-specific syntax, conventions, or mental models.} For instance: \textit{“Hi guys. I am completely new to C++ and learning it at the moment. [...]
This is what I wrote, but it just does not want to take [SIZE + 1] as the size of the new array.”}
\change{This suggests that transitions into new languages are particularly fraught, as novices must build fresh mental models without yet having mastered foundational programming concepts.}

\smallskip
\noindent \underline{\textbf{Coding Curiosity and Ambition}} (14 instances): Illustrates strong desire and interest to learn programming and improve skills in it. For example:
\textit{“I have curiosity of how other programs work, i like to explore cool projects, sometimes read their source code, and sometimes challenge myself to re-implement or improve it [...]”}.
\change{These examples demonstrate how curiosity can function as a positive emotional driver, motivating learners to persist despite obstacles.}

\smallskip
\noindent \underline{\textbf{\change{Programming Language} Learning Enthusiasm}} (14 instances): Focuses on individuals passionate about gaining knowledge and finding resources to master foundational programming languages, such as C, C++. \change{This finding is consistent with stack overflow survey on popular beginner language~\cite{stackoverflow2024survey}.} For example:
\textit{“I’m CS student and I had studied Basics of C++ and one course of data structure using C++ and actually I liked C++ and I want to get dipper in it what I should do ,in what I can use this language ??And can you give me a website to practice it ?”}
\change{This example indicates how enthusiasm for particular languages can energize learning trajectories, but also generate uncertainty about how best to channel that motivation into effective practice.}

\begin{figure*}[tb]
    \centering
    \includegraphics[width=0.8\textwidth]{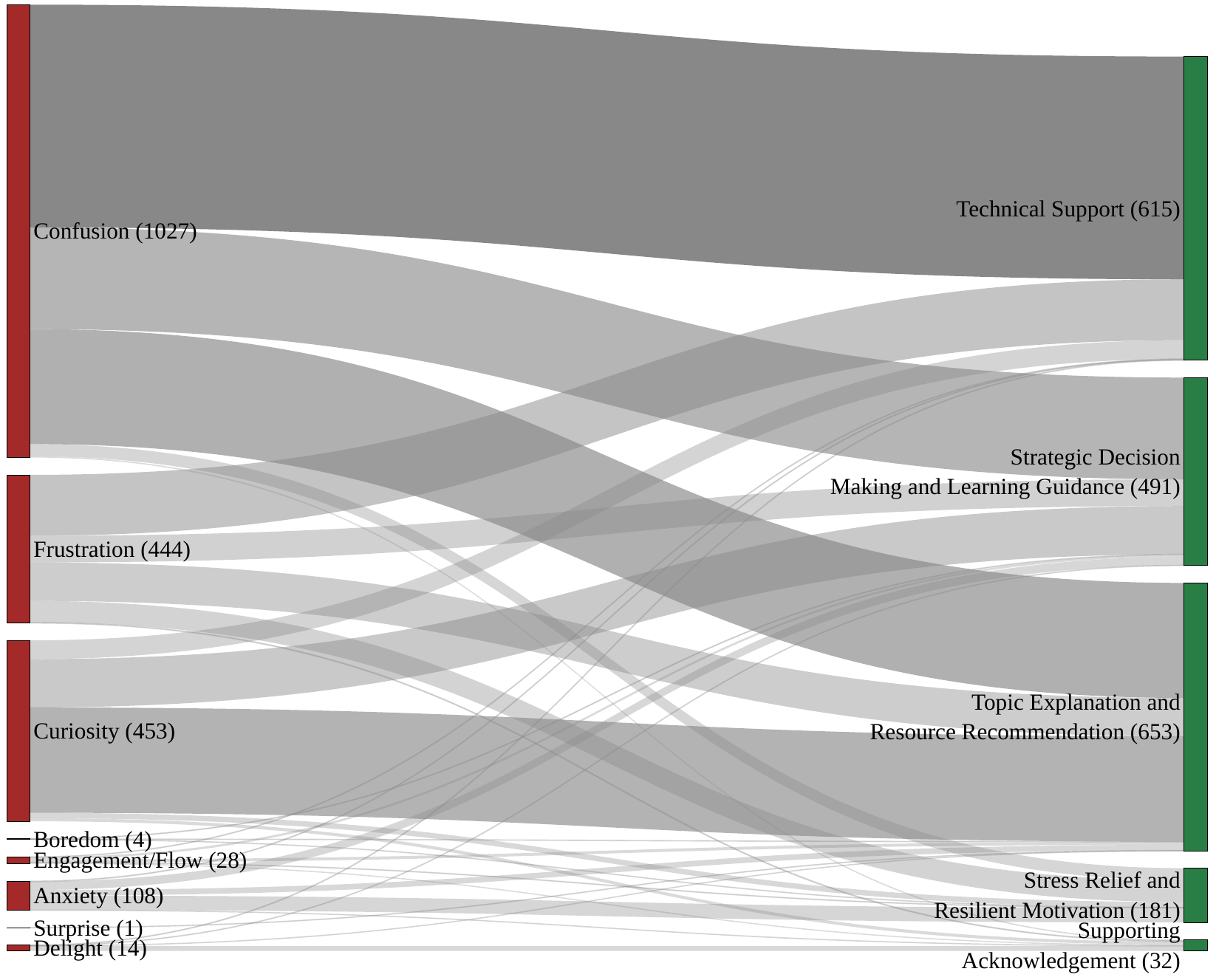}
    \caption{Mapping Novice Programmers’ Emotions to Support Needs (note: this is many-to-many connection).}
    \label{fig:emotion_to_support_category_sankey}
\end{figure*}

\section{RQ3: SUPPORT NEEDS OF NOVICE PROGRAMMERS}\label{system}

\textit{RQ3: What types of support do novice programmers need in the r/learnprogramming community?}

\subsection{Annotation}

As described in Section~\ref{dataset}, we performed open coding and axial coding~\cite{charmaz2006constructing} to understand what types of support novice programmers explicitly or implicitly request in their \textit{r/learnprogramming} posts. 
The annotators identified five key areas where novice programmers need support in informal space: \textit{Stress Relief and Resilient Motivation}, \textit{Topic Explanation and Resource Recommendation}, \textit{Strategic Decision Making and Learning Guidance}, \textit{Technical Support}, and \textit{Supporting Acknowledgment}.
Since 473 posts contained multiple support areas, we ended up with a total of 1,973 total count in 1,500 posts.

\subsection{Findings}
Figure~\ref{fig:emotion_to_support_category_sankey} illustrates the mapping between novice programmers’ emotional states and the categories of support they most frequently request, showing how affective struggles such as \textit{Confusion} or \textit{Frustration} are directly tied to specific support needs. Below we discuss the identified five categories of support needs that novice programmers express in \textit{r/learnprogramming}:

\smallskip
\noindent \underline{\textbf{Topic Explanation \& Resource Recommendation}} (653 instances):
This emerged as the most frequent need. Learners often explicitly requested better examples, alternative metaphors, and contextual explanations tailored to their current level of understanding, directly connecting to our RQ1 finding that \textit{Confusion} dominates emotional experiences. Emotion analysis of these posts confirms this pattern: \textit{Confusion} (374 instances) and \textit{Curiosity} (341 instances) are the most prevalent emotions, followed by \textit{Frustration} (181) and \textit{Anxiety} (40). These emotional signals indicate that users are not only lost but also eager to understand and improve. For example, the following post seeks help on loops, ``\textit{i am extremely bad at for loops and i want to improve but i cannot find any good source or site to practice them [...]}''. In such cases, well-structured explanations and resource recommendations are not just helpful, they are essential.

\smallskip
\noindent \underline{\textbf{Technical Support}} (615 instances):
This category captured learners’ struggles with programming tools, environment setup, and debugging tasks. Posts often involved issues such as installation errors, compiler messages, or unexpected runtime behavior. Emotion analysis shows that these experiences are strongly associated with \textit{Confusion} (508 instances) and \textit{Frustration} (201), followed by \textit{Curiosity} (62). These emotions reflect not just momentary setbacks but persistent blocks to progress. One learner describes: “\textit{What am I doing wring? I a completely new to programming and coding. I picked up a book for beginners yesterday and have been practicing the very basics. I am stuck at the section on Looping. [...] But when I try to execute the code, I get "SyntaxError: invalid syntax" with the p in print("Go!") highlighted. [...] 
It is driving me crazy!! [...]}” Such cases highlight the need for timely, specific, and reassuring technical support to sustain learner motivation and continuity.

\smallskip
\noindent \underline{\textbf{Strategic Decision Making \& Learning Guidance}} (491 instances):
This category reflects learners seeking direction, what to study, in what order, using which tools or languages. Emotion analysis highlights a mix of \textit{Confusion} (334 instances), \textit{Curiosity} (206), and \textit{Frustration} (148), along with a notable presence of \textit{Anxiety} (63). These emotions signal both uncertainty about the learning journey and a strong desire to make informed, effective decisions. One user shares: “\textit{How to learn JavaScript in the easiest, quickest, and most beneficial way?... I end up failing because I do not know what to do... I feel as if I am digging myself a hole and don’t know what to do but to give up.}” This illustrates the emotional weight behind strategic decisions and the importance of personalized, motivating guidance to prevent dropout and enhance self-efficacy. Recommending learning path to scaffold learner planning and goal setting can be beneficial to novices~\cite{yan2024generative}.  

\smallskip
\noindent \underline{\textbf{Stress Relief \& Resilient Motivation}} (182 instances):
This category includes posts where learners express burnout, self-doubt, or emotional fatigue related to the demands of learning to code. Emotion analysis highlights high levels of \textit{Frustration} (129 instances), \textit{Anxiety} (91), and \textit{Confusion} (90) -- indicating psychological strain and a strong need for emotional regulation and motivational support. One student shares: “\textit{I'm doing a CS degree and I'm only struggling in the coding classes... Java is the most difficult thing I have ever done in my life [...] I’ve never struggled so badly.}” Such expressions reflect the compounded effect of technical difficulty and emotional stress. In such scenarios, monitoring emotional state and delivering timely motivational or regulatory interventions can be helpful~\cite{wang2024synergy}.

\smallskip
\noindent \underline{\textbf{Supporting acknowledgment}} (32 instances):
This category includes posts where learners sought feedback, encouragement, or recognition, rather than direct help with technical problems. Although relatively infrequent, these posts were marked by positive affect and social intent. Emotion analysis shows \textit{Curiosity} (19 instances) and \textit{Delight} (14) leading. These emotions indicate a desire for validation, social learning, and reflective growth. One user writes: “\textit{I'm currently studying Java... I put in mind to become the best at Java. Where can I possibly invest in it in the future? I'd like to hear experiences of Java experts.}” Acknowledging such efforts reinforces confidence and fosters a sense of belonging, which are critical for long-term learner retention and identity development. Providing lightweight motivational feedback mechanisms, such as badges or affirming messages, to reinforce learner persistence and celebrate milestones is helpful~\cite{besser2020impact}.

\section{IMPLICATIONS AND RECOMMENDATIONS}\label{implications}

We outline the practical and research implications derived from our analysis, and propose a set of intelligent support needs to meet the affective and cognitive necessity of the novice learners in informal learning environments.

\smallskip
\noindent \underline{\textbf{Implications for Practitioners}}:
Our findings offer several practical considerations for those designing tools and environments that support novice programmers in informal settings. Practitioners (e.g., participants, moderators, and educational concent creators) of online communities can benefit from incorporating affect-sensitive and learner-centered design elements.

First, tools that can detect expressions of \textit{Confusion} or \textit{Frustration} in user-generated content may help flag posts needing timely intervention. Automated tagging systems can guide moderators or peers to respond more effectively~\cite{milne2019improving}. Second, recommendation systems that adapt to learners’ emotional tone and stated needs could better align resources with their current state. For example, when a learner expresses \textit{Confusion}, directing them to simplified explanations, adaptive tutorial-personalized to the user needs, or peer-validated tutorials may reduce cognitive overload~\cite{d2014confusion, becker2023programming}. Third, debugging remains a major source of emotional struggle. Practitioners might consider embedding diagnostic aids, such as contextual hints or example-driven walkthrough, that offer incremental support while preserving learner autonomy~\cite{becker2023programming}. Finally, positive reinforcement mechanisms, such as simple recognition of milestones, can help foster motivation~\cite{besser2020impact}. These do not need to be elaborate, automated acknowledgments or community badges tied to learner-declared achievements may suffice.


\smallskip
\noindent \underline{\textbf{Implications for Researchers}}:
Researchers should develop models and analytics that account for the fragmented, dynamic, and emotionally charged nature of learner interactions in informal spaces. Existing affective frameworks may require adaptation to reflect the informal discourse styles and episodic nature of help-seeking behavior on platforms like Reddit~\cite{haythornthwaite2018learning}. Study on adaptive learning~\cite{farahani2024towards, becker2023programming} and AI-assisted pair programming~\cite{bird2022taking} can be helpful in this regard. Longitudinal studies tracking learner progression across time and platforms can reveal how emotional states and support needs evolve, while comparative analyses can examine how platform features influence learner outcomes. Finally, there is a pressing need to design and evaluate lightweight, scalable interventions, such as emotion-aware bots, that are suited to the informal, asynchronous, and large-scale nature of these communities.


\begin{table*}[t]
    \centering
    \caption{Proposed Affect-Aware Support Bots for Novice Programmers}
    \begin{tabularx}{\textwidth}{|l|X|X|X|}
        \hline
        \textbf{Bot Name} & \textbf{Primary Function} & \textbf{Trigger Examples} & \textbf{Contextual Intervention} \\
        \hline
        ConceptBot & Explains programming topics & ``I don't get loops'', ``What is recursion?'' & Layered explanations, beginner-friendly resources, tutorial links \\ \hline
        PathBot & Guides learning paths & ``What should I learn next?'', ``Career roadmap?'' & Topic sequences, learning roadmaps, community wiki links \\ \hline
        DebugBot & Provides technical debugging help & ``My loop runs infinitely'' & Step-by-step debugging, topic sequences, community wiki links \\ \hline
        SupportBot & Addresses emotional challenges & ``Programming is hard'', ``I'm too dumb for this'' & Empathetic validation, coping strategies, peer success stories, integration with ConceptBot \\ \hline
        CelebrationBot & Reinforces achievements & ``I built my first app!'', ``Solved the bug!'' & Positive reinforcement, digital badges \\
        \hline
    \end{tabularx}
    \label{tab:supportbots}
\end{table*}

\noindent \underline{\textbf{Designing Support Bots for Novice Programmers}}:
To operationalize our findings, we propose five automated intelligent  affect-aware bots aligned with the key support needs identified in our study. These bots  will function by responding to textual cues with context-sensitive interventions targeting specific emotional and cognitive challenges. As Reddit supports automated bots, these can be integrated in any specific subreddit needs within current Reddit's infrastructure. Table~\ref{tab:supportbots} summarizes the proposed bots, their primary functions, typical triggers, and forms of contextual intervention. Below we discuss them.

\noindent
\textbf{ConceptBot} will address topic explanations by detecting semantic phrases like ``\textit{I don't get loops}'', and respond with layered adaptive explanations and beginner-appropriate resources and thus reducing confusion by addressing common learning difficulties~\cite{prather2023robots, bozkurt2024manifesto, becker2023programming}. 

\noindent
\textbf{PathBot} will support strategic decision-making, activated by expressions like ``\textit{What should I learn next?}''. It will suggest learning pathways aligned with learner goals, pointer community wiki, suggest topic sequences or alternative paths learning~\cite{yan2024generative}.

\noindent
\textbf{DebugBot} will provide technical assistance by parsing error-related queries like ``\textit{my loop runs infinitely.}'' It will analyze code snippets, identify error patterns, and offer step-by-step debugging advice. When issues reflect conceptual misunderstanding, DebugBot integrates with ConceptBot. Generative AI can be helpful here~\cite{becker2023programming}. AI-assisted pair programming, already an area of active research, can be leveraged here~\cite{bird2022taking, zakaria2022two}.

\noindent
\textbf{SupportBot} will address affective challenges when learners express distress through phrases like ``\textit{I'm too dumb for this.}''. It will respond with empathetic validation and suggest coping strategies such as micro-goals, or peer success stories~\cite{wang2024synergy}.

\noindent
\textbf{CelebrationBot} will promote positive reinforcement by recognizing achievements in posts like ``\textit{I built my first app!}''~\cite{besser2020impact}. 
\section{THREATS TO VALIDITY}

We acknowledge several key threats to validity in our study. Below we describe them:

\subsubsection*{Construct Validity}
We adopted the LCE framework to annotate emotional states. While this framework is grounded in learning science, it may not fully capture culturally specific or domain-specific emotional expressions that arise in informal programming contexts. Additionally, Reddit posts represent episodic, text-based expressions of emotion, often lacking context about learning duration, personal background, or environment. As such, our measurement of affect and learning needs are necessarily partial and inferred from language, which may deviate from actual learner experiences.

\subsubsection*{Internal Validity}
Our emotion identification methodology relies on manual annotation, can potentially introduce subjectivity despite substantial inter-rater agreement and subsequent complete agreement~\cite{imtiaz2018}. 
Secondly, our initial filtering of posts using the \textit{Llama} model-based on heuristic prompts for emotional content and novice identity, could introduce bias if the model misclassified posts. To reduce these risks, we employed a multi-stage filtering process (\textit{Llama} + human verification), used a theoretically grounded annotation schema, and conducted reliability checks.  
Another concern stems from filtering out posts labeled as neutral by LLMs, which may inadvertently exclude emotionally nuanced posts that are subtle or indirect in their expression. While this trade-off may result in the loss of some valid emotional data, we believe the selected dataset, comprised of 1,500 manually verified, emotionally expressive posts, provides sufficient depth and diversity to support robust analysis. This approach aligns with prior work~\cite{imran2022data} and follows recommended domain practices for dataset curation~\cite{koufakou2025review}.

\subsubsection*{External Validity}
The dataset is limited to posts from Reddit’s \textit{r/learnprogramming} community. This sample may not generalize to learners in formal educational settings or on other platforms. Reddit users may differ in demographics, communication style, and goals compared to students in courses or participants in mentoring systems.
As a self-selected help-seeking community~\cite{cross2017classifying}, it may overrepresent emotionally negative experiences and underrepresent silent, successful learners. Thus, while offering insight into public learning struggles, the data may not reflect the full range of emotional experiences. Broader validation across diverse platforms is needed.

\section{CONCLUSION AND FUTURE WORK}
This study examined the emotional experiences of novice programmers in informal online environments through a large-scale analysis of \textit{r/learnprogramming} posts. Using the Learning-Centered Emotions framework, we identified \textit{Confusion} and \textit{Frustration} as dominant emotional states, often linked to unclear learning paths, debugging difficulties, and resource misalignment. Positive emotions were rare but revealed important moments of progress and motivation. Based on these findings, we proposed five categories of learner support needs and outlined a design vision for lightweight, affect-sensitive bots targeting explanation, guidance, debugging, emotional support, and achievement acknowledgment.

Our future work will focus on developing and deploying the feasibility of these support systems for the proposed bots, with particular attention to their usability and impact in live community settings. Additionally, we plan to explore how learner trajectories evolve over time and whether targeted interventions improve persistence and engagement. \change{While our dataset is drawn from Reddit, many of the identified challenges, such as emotional overwhelm, lack of structured guidance, and the need for peer validation, may also be relevant across other informal learning platforms like Stack Overflow, Discord, or YouTube. These platforms differ in interaction norms, anonymity, and moderation, which may shape how emotions are expressed or addressed.} Future research should examine these variations to assess the broader applicability of our findings and guide the design of platform-specific support strategies.

\balance
\bibliographystyle{ACM-Reference-Format}  
\bibliography{references}


\end{document}